\documentclass[aip,apl,
 amsmath,amssymb,
 reprint,%
]{revtex4-1}

\usepackage{graphicx}
\usepackage{dcolumn}
\usepackage{bm}

\usepackage[utf8]{inputenc}
\usepackage[T1]{fontenc}
\usepackage{mathptmx}
\usepackage{etoolbox}

\usepackage{xcolor}
\usepackage{float}
\usepackage{amsmath}
\usepackage{amssymb}
\usepackage{dcolumn}
\usepackage{url}

\makeatletter
\def\@email#1#2{%
 \endgroup
 \patchcmd{\titleblock@produce}
  {\frontmatter@RRAPformat}
  {\frontmatter@RRAPformat{\produce@RRAP{*#1\href{mailto:#2}{#2}}}\frontmatter@RRAPformat}
  {}{}
}%
\makeatother
\begin{document}


\title[A tunable photonic band gap resonator for axion dark matter searches]{A tunable photonic band gap resonator for axion dark matter searches}
\author{Samantha M. Lewis}
 \affiliation{Department of Physics and Astronomy, Wellesley College, Wellesley, Massachusetts 02481, USA\looseness=-1}
  \email{sl128@wellesley.edu}
\author{Dillon T. Goulart}
\altaffiliation{Now at University of California, Davis}
\affiliation{Department of Nuclear Engineering, University of California, Berkeley, Berkeley, California 94720, USA\looseness=-1}
\author{Mirelys Carcana Barbosa}
\altaffiliation{Now at Brown University}
\affiliation{Department of Nuclear Engineering, University of California, Berkeley, Berkeley, California 94720, USA\looseness=-1}
\author{Shriram Jois}
\affiliation{Department of Nuclear Engineering, University of California, Berkeley, Berkeley, California 94720, USA\looseness=-1}
\author{Alexander F. Leder}
\affiliation{Department of Nuclear Engineering, University of California, Berkeley, Berkeley, California 94720, USA\looseness=-1}
\affiliation{Physics Division, Los Alamos National Laboratory, Los Alamos, New Mexico 87545, USA\looseness=-1}
\author{Aarav M. Sindhwad}
\altaffiliation{Now at Yale University}
\author{Isabella Urdinaran}
\author{Karl van Bibber}
\affiliation{Department of Nuclear Engineering, University of California, Berkeley, Berkeley, California 94720, USA\looseness=-1}

\date{\today}

\begin{abstract}
Axions are a well-motivated dark matter candidate particle. Haloscopes aim to detect axions in the galactic halo by measuring the photon signal resulting from axions interacting with a strong magnetic field. Existing haloscopes are primarily targeting axion masses which produce microwave-range photons and rely on microwave resonators to enhance the signal power. Only a limited subset of resonator modes are useful for this process, and current cylindrical-style cavities suffer from mode mixing and crowding from other fundamental modes. The majority of these modes can be eliminated by using photonic band gap (PBG) resonators. The band gap behavior of these structures allows for a resonator with mode selectivity based on frequency. We present results from the first tunable PBG resonator, a proof-of-concept design with a footprint compatible with axion haloscopes. We have thoroughly characterized the tuning range of two versions of the structure and report the successful confinement of the operating TM$_{010}$ (transverse magnetic) mode and the elimination of all transverse electric modes within the tuning range.
\end{abstract}

\maketitle

\section{Introduction}
The axion is a hypothetical particle that arises from a proposed solution to the Strong CP problem.\cite{PQ96,PQ97,FW78,SW78} Its properties also make the axion a well-motivated dark matter candidate particle.\cite{Preskill,FCDreview} One of the leading techniques to search for axions aims to measure the axion conversion to a photon in the presence of an applied magnetic field.\cite{SikivieHaloscope} In this process, the resulting photon carries the full energy of the axion. In the case of experiments searching for virialized axions in the galactic dark matter halo (called `haloscopes'), the photon frequency is related to the axion mass as
\begin{equation}
h\nu = m_ac^2\left[1+\frac{1}{2}\mathcal{O}\left(\beta^2\right)\right],
\label{eq:energy}
\end{equation}
where $\nu$ is the photon frequency, $h$ is Planck's constant, $m_a$ is the axion mass, $c$ is the speed of light, and $\beta \approx 10^{-3}$ is the galactic virial velocity.\cite{Graham}

Traditional axion haloscopes combine a high field magnet, dilution refrigerator, low noise readout system, and tunable electromagnetic cavity to search for the converted photon signal. The goal is to tune the frequency of a cavity mode to match the frequency of the converted photon, resulting in resonant enhancement of the signal. The efficiency of this conversion is highest when the cavity electric field is aligned with the applied magnetic field, quantified by a mode-dependent form factor $C_{nm\ell}$. For experiments using a solenoid with a uniform magnetic field along $\hat{z}$, the form factor is
\begin{equation}
C_{nm\ell} = \frac{\left(\int{d^3\mathbf{x}\,\mathbf{\hat{z}}\cdot\mathbf{E}_{nm\ell}\left(\mathbf{x}\right)}\right)^2}{V\int{d^3\mathbf{x}\,\epsilon\left(\mathbf{x}\right)\left|\mathbf{E}_{nm\ell}\right|^2}}
\label{eq:C}
\end{equation}
where $V$ is the cavity interior volume, $\epsilon$ is the relative permittivity of any material inside the cavity, and $\mathbf{E}_{nm\ell}$ is the electric field of the mode specified by indices $n$, $m$, and $\ell$. In traditional haloscope cylindrical cavities, the lowest order transverse magnetic (TM) mode, TM$_{010}$, has the highest possible form factor with an $E$~field purely along the $z$ axis. The overall cavity performance contributes to the signal power as
\begin{equation}
P \propto VQ_LC_{nm\ell}
\label{eq:power},
\end{equation}
where $Q_L$ is the loaded quality factor of the cavity mode.

Both the mass of the axion and the coupling strength of the axion to photon conversion $g_{a\gamma\gamma}$ are unknown. Various astrophysical constraints limit the upper bound of QCD cold dark matter axion masses to roughly 15~meV or 50~meV, depending on the model.\cite{Carenza:2019pxu,Snowmass,AxionLimits} The possible range covers many orders of magnitude in mass and therefore frequency. As a result, a variety of experiments are being performed and built around the world to cover this wide range, with search frequencies down to the kHz scale and up the THz scale.\cite{Snowmass} In the microwave frequency range, haloscopes use tunable cavities in order to search over a range of axion masses in each experimental run. The expected signal power is extremely small, so experiments must integrate for some time at each frequency point to achieve sensitivities corresponding to the model bands of interest for the QCD axion.

In traditional haloscope cavities, frequency tuning is achieved using one or more cylindrical tuning rods which can be rotated, changing the frequency of all TM modes. This method of tuning does not appreciably tune transverse electric (TE) and transverse electromagnetic (TEM) modes. For a standard cylindrical cavity with a metallic tuning rod, all possible TE, TM, and TEM modes are supported. As the operating TM$_{010}$ mode is tuned, it crosses the stationary TE and TEM modes. By definition, TE and TEM modes have no significant $E_z$ component and therefore have $C_{nm\ell}$ values of zero. As the TM$_{010}$ approaches a stationary mode, the two modes mix, reducing the $E_z$ component of the field and by extension the form factor of the hybridized mode. The reduction in form factor is significant over a range of frequencies around the stationary mode, making it unfeasible to search for axions in that range.\cite{CCCRSI} Such mode crossings are one cause for the gaps in published haloscope exclusion limits.\cite{ADMXresult,HAYSTACresult,CAPPresult} Filling in these gaps requires time-intensive mechanical reconfiguration of the cavity and more run time at the affected frequency range to achieve the same sensitivity.

Eliminating unwanted stationary modes would speed up axion searches by making more of the cavity tuning range usable in a given experiment, removing the need for cavity reconfiguration. Many of the cavity innovations of interest for next generation axion searches also face challenges with mode crossings or mode mixing.\cite{QUAX,RSI7rod} The mode crossing problem also limits the feasibility of using higher order TM modes for searches, or longer cylindrical cavities, where in both cases the density of TE modes is higher. Mode selectivity can be achieved using advanced resonator techniques such as microwave photonic band gap (PBG) structures.

In this paper, we present results from a proof-of-concept tunable PBG resonator designed to be used in an axion haloscope. Section \ref{sec:PBG} describes the principle of PBG structures and outlines the tunable resonator design. Section \ref{sec:meas} describes the methods used to characterize the device performance. Sections \ref{sec:4row} and \ref{sec:3row} detail the results of two test cases. The conclusions and future work are summarized in Section \ref{sec:summary}.

\section{Photonic band gap structure design}\label{sec:PBG}
\begin{figure}[!htb]
\centering
\includegraphics[width=0.6\columnwidth]{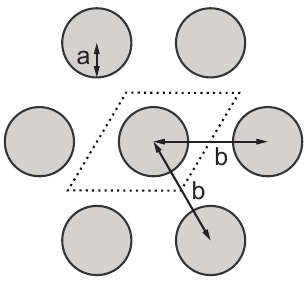}
\caption{Example 2D arrangement based on a triangular lattice, showing the rod spacing $b$ and radius $a$. The dashed line indicates the unit cell.}
\label{fig:base}
\end{figure}

A PBG consists of a periodic array of elements which give rise to a photonic band structure. For some configurations, there are band gaps in which photons of certain frequencies cannot propagate through the lattice.\cite{Yablonovitch93} These band gaps are analogous to electronic band gaps in materials.\cite{Yablonovitch1990} One type of 2D PBG lattice in the microwave range can be made using metallic cylindrical rods. Figure \ref{fig:base} shows the geometry of a triangular 2D lattice. The band structure is determined by the rod radius $a$ and the center-to-center rod spacing $b$.\cite{Smirnova2002,SmirnovaThesis}

A defect can be created in a PBG lattice to confine a mode with a frequency within the band gap. Confinement occurs even without an overall outer reflecting boundary, so any excited modes with frequencies outside of the band gaps can propagate out of the structure. While PBG theory is based on an infinite lattice, experiments have shown as few as 3 rows of rods can give sufficient confinement of the desired mode.\cite{SmirnovaThesis} Many 2D PBG structures have been successfully used to confine a variety of chosen TE and TM modes in waveguides and resonators.\cite{MunroePBG,SimakovPBG,SirigiriPBG,ZhangPBG,NanniPBG,ArsenyevPBG}

A triangular, metallic PBG lattice has several large band gaps for TM modes, but very few band gaps for TE modes.\cite{Smirnova2002} As a result, a variety of lattice configurations will support TM modes with no nearby TE modes, the ideal scenario for axion haloscopes.

The previously cited microwave PBG resonators and waveguides are all designed to operate at a single precise frequency. In order to use a PBG lattice as a resonator for an axion search, it must be possible to freely tune the mode confined in the defect over a wide frequency range without reintroducing TE modes. To test whether it is possible to break symmetry in the structure and maintain the band gap behavior, we designed a resonator for an axion haloscope using the same tuning method as current cavities: an off-axis, cylindrical tuning rod which tunes a TM$_{010}$-like mode. Design simulations were performed using both CST Microwave Studio\cite{CST} and Ansys HFSS.\cite{HFSS} The structure was designed to have $a/b=0.43$ with $a=0.125~\text{in}$, corresponding to a rod spacing of $b=0.2907~\text{in}$. The height of the lattice between the endcaps is 10~in. A tuning with radius $r=0.32~\text{in}\approx1.1b$ gives a tuning range of roughly 7.352--9.343~GHz. The size of this tuning range is comparable to similar existing haloscope cavities.

\begin{figure}[htb]
\centering
\includegraphics[width=\columnwidth]{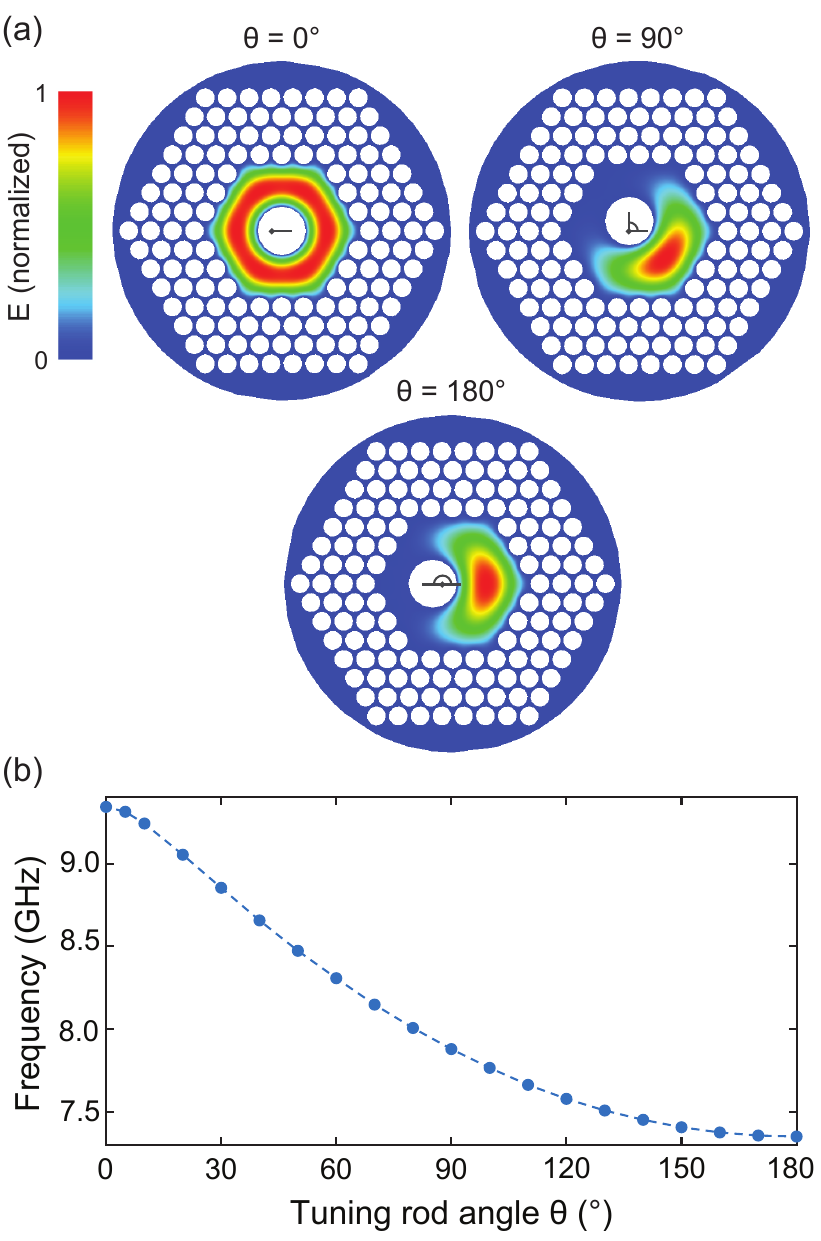}
\caption{(a) Normalized magnitude of the electric field in the structure at tuning rod angles of $\theta=0^{\circ},~90^{\circ},~\text{and}~180^{\circ}$. The HFSS simulation shown here includes a radiating boundary condition on the outer cylindrical surface. (b) Tuning curve generated from HFSS simulations of the structure including antenna coupling and the alignment grid. Simulations of $10^{\circ}$ steps are shown, with an additional point at $5^{\circ}$ to show the curvature of the tuning at the highest frequencies.}
\label{fig:4rowtuning}
\end{figure}

Figure \ref{fig:4rowtuning} shows a top-down view of the electric field in the lattice at three tuning positions and the corresponding simulated tuning curve of the TM$_{010}$-like mode. The simulated performance of this structure is comparable to that of existing copper cylindrical cavities, with room temperature $Q_0$ values of order $10^{4}$ for copper rods and endcaps. The structure as designed supports TM modes and no TE modes within the tuning range. TEM modes are introduced due to the metal tuning rod. However, previous work has shown that TEM mode crossings, while undesirable, have a smaller frequency range over which they reduce the form factor, making them more tolerable.\cite{CCCRSI}

The PBG structure was built using solid copper lattice rods, a copper tuning rod, and 1~in thick copper endcaps. Each PBG rod is press-fit into pockets in the endcaps. Six rods at the outer corners of the structure are tapped on both ends to bolt the structure together. The length of the tuning rod is 9.984~in, designed to leave 0.008~in gaps on either side of the tuning rod for rotation. A 1~mm thick PETG plastic 3D-printed alignment grid is used to hold the rods in place during assembly. This alignment grid was included in the simulation results presented and has minimal impact on the electrical performance.

\section{Measurement goals and methodology}\label{sec:meas}
Our goal with this proof-of-concept structure was to determine whether the introduction of the tuning rod (breaking symmetry in the defect) affects the band gap behavior or causes the re-introduction of TE modes. While frequency tuning behavior can be determined solely from transmission measurements of the cavity, bead-pull measurements are necessary to verify that the mode being tuned is the TM$_{010}$ mode, study mode crossings, and ensure there is no unexpected mixing from any hard-to-see modes. 

\begin{figure}[!htb]
\centering
\includegraphics[width=\columnwidth]{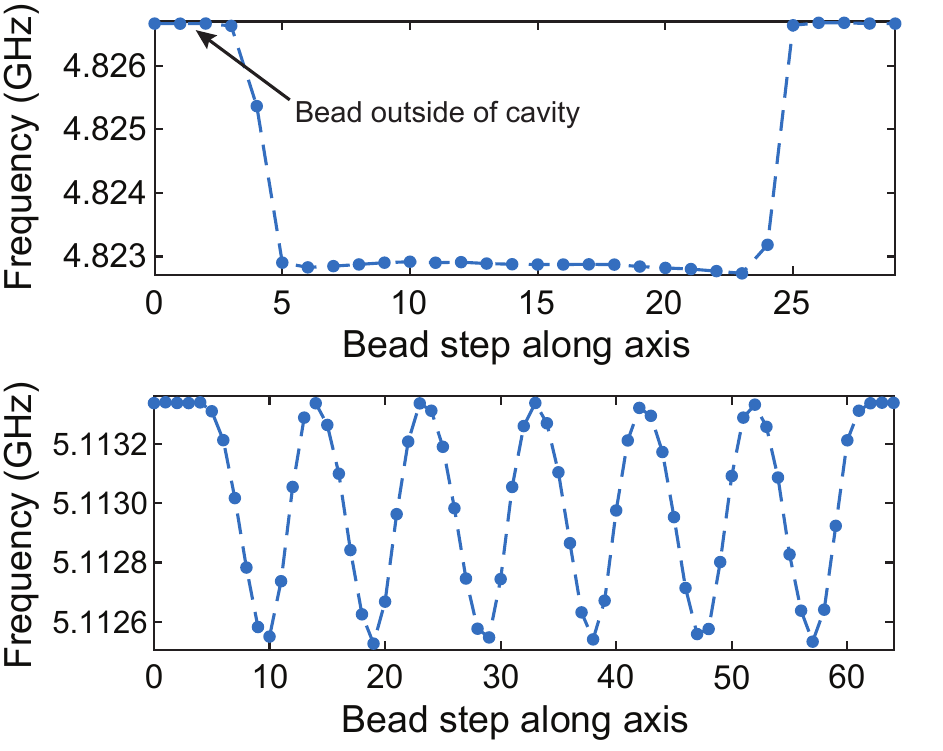}
\caption{Example bead-pull measurements from two different cavities. In both cases, the frequency is constant at the unperturbed value when the bead is outside of the cavity. The top shows a near ideal TM$_{010}$, with a constant field along $z$ resulting in a near constant perturbed frequency. This measurement was taken from an early, shorter prototype of the PBG with no tuning rod. The bottom shows an example TE mode from a standard cylindrical cavity, where the frequency returns to the unperturbed value at nodes in the $E$ field.}
\label{fig:beadpullexample}
\end{figure}

Bead-pull measurements map out the field profile in one dimension by introducing a small perturbation that shifts the mode frequency by an amount proportional to the square of the local electric field.\cite{BP1,BP2} An ideal TM$_{010}$ mode has a constant electric field in the longitudinal direction, giving a constant frequency versus bead position. All TE and TEM modes have nodes in the longitudinal direction, resulting in oscillations of the frequency as a function of bead position. Figure \ref{fig:beadpullexample} shows example bead-pull measurements for a TM$_{010}$ mode and a TE mode.

\begin{figure}[!htb]
\centering
\includegraphics[width=\columnwidth]{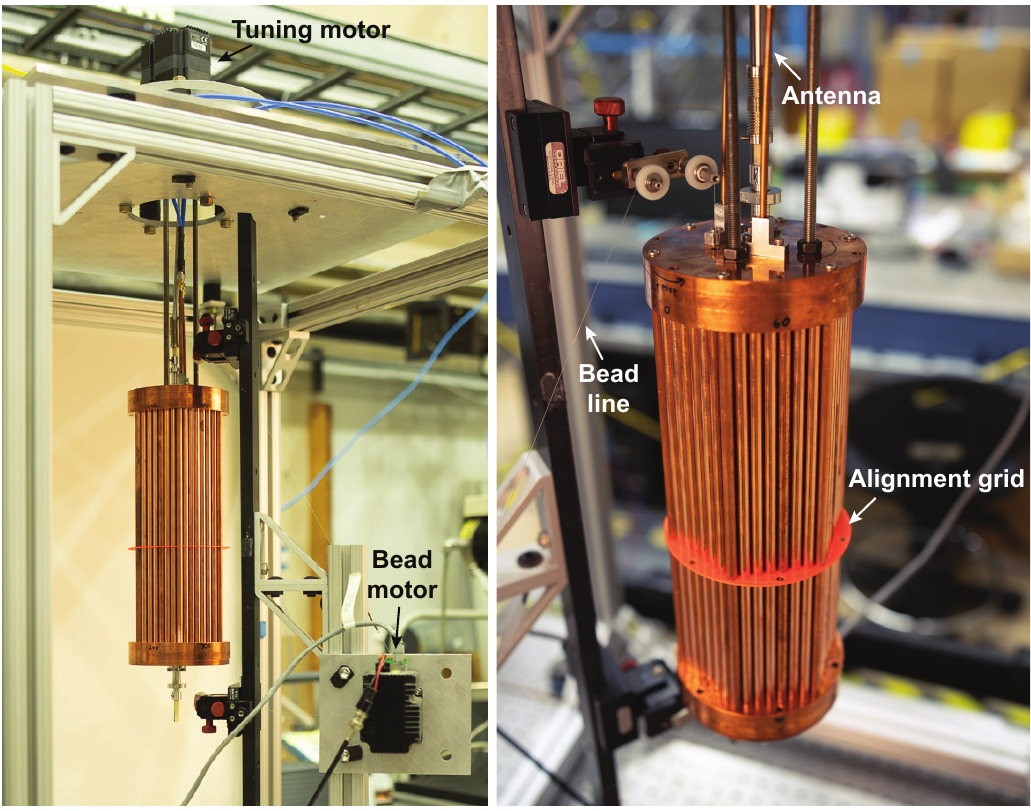}
\caption{The assembled PBG structure mounted on the bead-pull test stand. The left photo shows both motors used in the measurements. The right photo provides a close-up view showing the bead line, one variable coupling antenna, and the alignment grid.}
\label{fig:photos}
\end{figure}

The mode frequency is determined by measuring the peak in the transmission spectrum ($S_{21}$) using a vector network analyzer (VNA). Our test stand uses a Keysight E5071C ENA. We couple to the cavity using two adjustable coaxial probe antennas located at the top of the cavity. The rotation of the tuning rod and the bead line are controlled using stepper motors. For our bead, we used a 1/8 inch diameter nylon sphere on monofilament nylon suture line.\cite{HAYSTACNIM,CCCRSI} Photos of the PBG structure mounted on the test stand are shown in Fig. \ref{fig:photos}.

To test the PBG performance and effectiveness at eliminating TE modes, we measured the TM$_{010}$ mode profile at $1^{\circ}$~steps in tuning rod angle throughout the entire tuning range. TE and TEM modes typically couple poorly to the antenna orientation used and may not be visible in a measured spectrum until they are mixing with a nearby TM mode. To ensure there were no unexpected TE modes that may be poorly confined with low $Q$ values or limited coupling, we repeated the measurement procedure at several antenna couplings.

Vibrations of the bead line, cavity, or tuning rod can cause changes to the bead-pull measurement. The test stand is built with vibration isolation between components, but additional wait times are occasionally needed for clean measurements. As part of this work, full automation of the bead-pull test stand was developed, allowing for longer wait times for bead stabilization. This upgrade also made it possible to sweep the tuning rod angle in fine steps over the entire tuning range.

\section{Four-row design results}\label{sec:4row}
\begin{figure}[!htb]
\centering
\includegraphics[width=1\columnwidth]{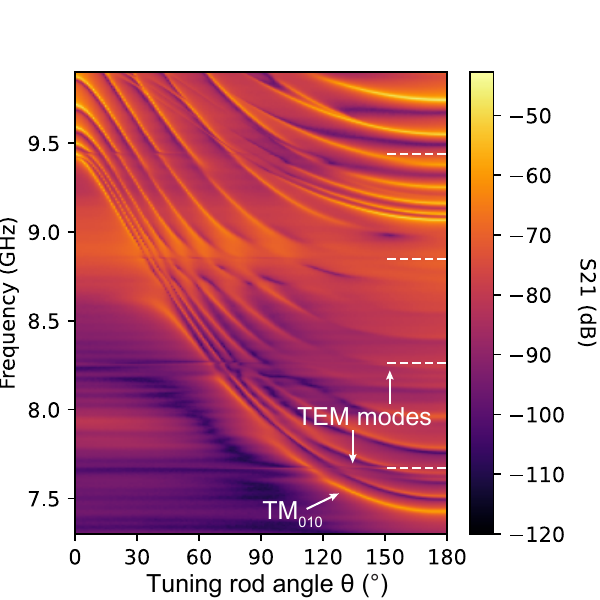}
\caption{Mode map for the tuning range of the 4~row PBG configuration. The color scale represents $S_{21}$ through the PBG, with modes visible as peaks. Dashed lines indicate the expected TEM mode frequencies.}
\label{fig:4rowmm}
\end{figure}

The design tuning range of the 4-row structure was 7.352--9.343~GHz (Fig. \ref{fig:4rowtuning}). The measured TM$_{010}$ mode range was slightly higher, from 7.428--9.409~GHz. There were a number of performance limitations in this prototype design. The measured mode map for the 4-row design is shown in Fig. \ref{fig:4rowmm}, where these regions are visible. In the measurement range, four TEM modes are expected at 7.67, 8.26, 8.85, and 9.44~GHz. Each is visible on the mode map, especially at angles where they are mixing with a nearby TM mode. In other regions, the TEM modes are less visible due to coupling to the antenna, as previously described. This limited visibility motivates the need for bead-pull measurements at all rod angles to ensure there is truly no mixing. No unexpected modes were encountered, but there were several performance limitations that reduced bead-pull measurement quality in portions of the tuning range.

Most notably, the electrical contact between the PBG rods and endcaps was variable, resulting in lower than expected $Q$ values at all frequencies. The aspect ratio of the defect (height compared to radius) was also more extreme than typical haloscope cavities. Since the antennas are located at the endcaps, high aspect ratio can lead to axial mode localization and reduced coupling to the TM$_{010}$ mode at some tuning angles. This same limitation has been observed in other high aspect ratio cylindrical test cavities. Additionally, two broadband features (visible in Fig. \ref{fig:4rowmm}) obscured the TM$_{010}$ mode near 7.8 and 8.6~GHz. In these regions, transmission between the two antennas was high overall with no clear resonances. The origin of these features is unknown, but may be due to the combination of the high aspect ratio and gaps between the tuning rod and the endcaps. Localized field near the ends of the cavity may have caused increased coupling of the field into the gaps, one of which is located near the antennas. Further modeling work is ongoing to explore the origins of these broadband features.

\section{Three-row test results}\label{sec:3row}
\begin{figure}[!htb]
\centering
\includegraphics[width=\columnwidth]{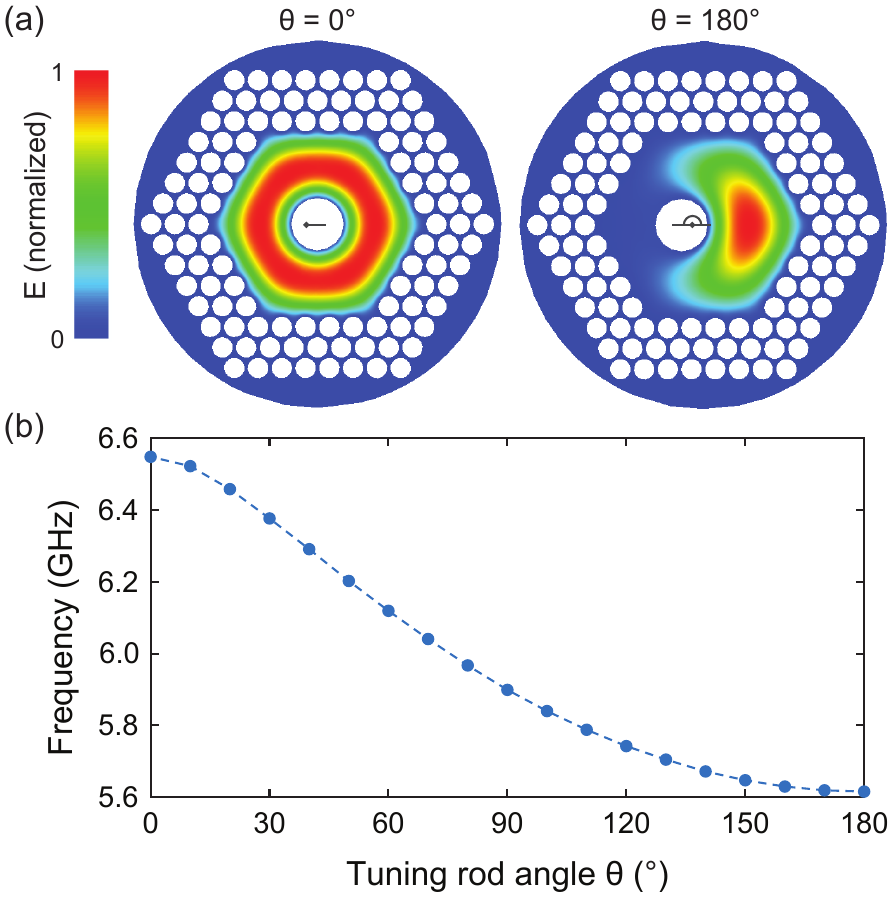}
\caption{(a) Normalized magnitude of the electric field in the structure at tuning rod angles of $\theta=0^{\circ}~\text{and}~180^{\circ}$. (b) Tuning curve generated from HFSS simulations of the structure including antenna coupling and the alignment grid. Simulations of $10^{\circ}$ steps are shown.}
\label{fig:3rowtuning}
\end{figure}

To provide better mode visibility over the entire range of rod angles and verify there were no unexpected modes, we tested a version of the design with the innermost row of the PBG lattice removed. This 3-row configuration increased the defect radius, reducing the aspect ratio and improving coupling significantly. The larger defect size shifts the measured tuning range to 5.659--6.567~GHz. This is similar to the simulated tuning range of 5.616--6.548~GHz. The full simulated tuning behavior is shown in Fig. \ref{fig:3rowtuning}.

\begin{figure}[!htb]
\centering
\includegraphics[width=1\columnwidth]{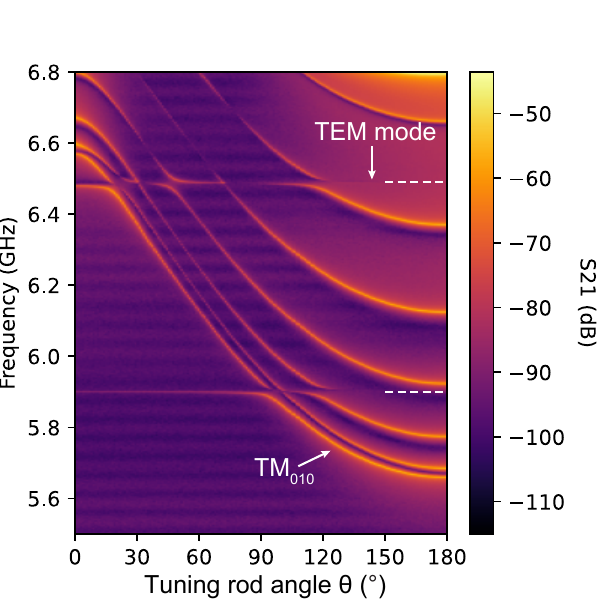}
\caption{Mode map for the tuning range of the 3~row PBG configuration. The color scale represents $S_{21}$ through the PBG, with modes visible as peaks. The only modes visible are the expected TM and TEM modes (indicated by the dashed lines). The ripple visible at very low $S_{21}$ is likely a standing wave in the measurement network not removed by calibration. This ripple is close to the noise floor of the VNA.}
\label{fig:3rowmm}
\end{figure}

We repeated the same procedure used to characterize the 4-row design, performing bead-pull measurements over the full tuning range in $1^{\circ}$~steps with multiple antenna couplings. A full mode map is shown in Fig. \ref{fig:3rowmm}. The improved coupling due to the reduced aspect ratio is apparent. The broadband features seen in the 4-row configuration do not appear in this larger radius structure; this supports the hypothesis that the high aspect ratio and poor coupling of the 4-row design was causing mode localization near the end of the structure. Two TEM modes are expected at 5.90 and 6.49~GHz, both of which are visible in the mode map. No unexpected modes were encountered in any part of the tuning range. This determination was made by observing the shape of the mode profile at every tuning angle with several antenna couplings.

\begin{figure}[!htb]
\centering
\includegraphics[width=\columnwidth]{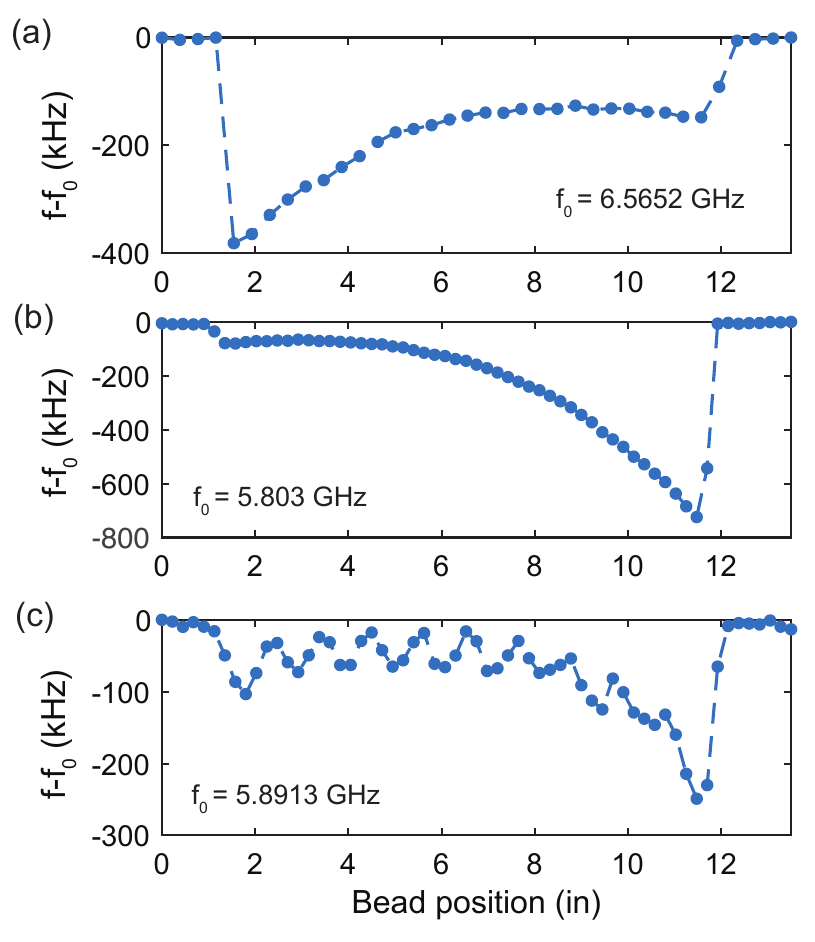}
\caption{Sample bead-pull measurements of the TM$_{010}$ mode at multiple rod angles: (a)~near the maximum of the tuning range, (b)~near the minimum of the tuning range, and (c)~in the mode crossing with the 10th order TEM mode.}
\label{fig:bp3row}
\end{figure}

Figure \ref{fig:bp3row} shows sample bead perturbation measurements of the TM$_{010}$ without and with mode mixing. In Fig. \ref{fig:bp3row}a and b, two typical profiles are shown near the maximum and minimum of the tuning range. The asymmetry of the mode from the bottom to the top of the cavity has been observed in previous cylindrical cavities using this tuning mechanism and changes with rod angle. The impact of this asymmetry on the form factor is negligible.\cite{CCCRSI} Figure \ref{fig:bp3row}c shows an example of the TM$_{010}$ mode mixing with the 10th order TEM mode at 5.90~GHz. The oscillation of the $E$ field of the hybridized mode is clearly visible.

\section{Summary and future work}\label{sec:summary}
We have performed a thorough study of the tuning performance of a tunable PBG resonator, designed to support TM modes and eliminate TE modes. The tuning mechanism and overall structure size were chosen to be compatible with current haloscope techniques. We have successfully demonstrated the elimination of all TE modes, even with a large symmetry-breaking perturbation introduced by a tuning rod. We have also demonstrated the robustness of the PBG design, showing that even 3 rows of the PBG lattice provides sufficient confinement of the TM$_{010}$ mode to see tuning over the full angle range. Because the scan rate in axion searches increases as the interaction volume squared, it can be advantageous to use as few rows as possible to provide a larger interaction volume for a given magnet bore size. Further work has been undertaken to explore the possibility of using only 2 rows of rods, showing good confinement over a wide portion of the tuning range.\cite{PBG7rod}

The $Q$ value of this cavity is limited to $\approx5\times10^3$ and does not achieve simulated room temperature values of~$\sim10^4$. The primary reason for this is poor electrical contact between the rods and the endcaps due to the rod-in-pocket design. As a design created to demonstrate the concept of a tunable PBG, the $Q$ value was not a performance metric and emphasis was placed on ease of fabrication rather than $Q$ value optimization. Work is ongoing to explore different fabrication methods to improve the $Q$ value to achieve at least parity with current copper cavities, such screwing or brazing all lattice rods to the endcaps. Results from a new structure where every rod is screwed to the endcaps has demonstrated significantly improved $Q$, on par with simulated expectations and standard copper cavity values.\cite{PBG7rod}

A PBG lattice boundary is compatible with other next-generation cavity innovations for axion searches. The TM$_{010}$ mode frequency is ultimately determined by the size of the inner region, so achieving higher frequencies requires additional complexity. With this successful demonstration of PBG behavior and the flexibility of the design, ongoing experiments are being performed to explore multi-rod tuning designs and the use of a PBG lattice around a metamaterial resonator, both of which are aimed at reaching higher frequencies. Results of a study based on a 7 tuning rod design with a 2 row PBG have been published, showing it is possible to incorporate a PBG boundary with more complex tuning mechanisms.\cite{PBG7rod} Further work will be performed to explore other materials and the use of higher order modes to improve $Q$ values and reach higher frequencies while maintaining large cavity volumes.\\

\section{Acknowledgements}
The authors thank Nolan Kowitt and S. Al Kenany for their work on the bead-pull test stand development, and AJ Gubser for the test stand photos. The authors also thank Sergio Velazquez and Amara Johnson for their work fabricating the copper structure. The authors gratefully acknowledge support from from the National Science Foundation under Grant No. PHY-2209556. M.C.B. gratefully acknowledges support from the Ronald E. McNair Scholars Program and the Semiconductor Research Corporation.
\bibliography{PBGfull}

\end{document}